\documentclass[aps,preprint]{revtex4}%
\usepackage{amsfonts}
\usepackage{amsmath}
\usepackage{amssymb}
\usepackage{graphicx}%
\begin{document}
\title[Short title for running header]{Collective Excitations Spectrum in Density Modulated One-Dimensional Electron
Gas in a Magnetic Field}
\author{K. Sabeeh, M. Tahir}
\affiliation{Department of Physics, Quaid-i-Azam University, Islamabad, Pakistan}

\pacs{73.20.Mf.; 71.45.Gm}

\begin{abstract}
We determine the collective excitations spectrum and discuss the numerical
results for a parabolically confined density modulated quasi-one dimensional
electron gas (1DEG) in the presence of an external magnetic field. We derive
the inter-and intra-band magnetoplasmon spectrum within the Self Consistent
Field (SCF) approach. In this work we focus on magnetoplasmon oscillations in
this system and as such results are presented for the intra-Landau-band
magnetoplasmon spectrum that exhibits oscillatory behavior, these oscillations
are not with constant period in $1/B$ and are significantly effected at low B
and corresponding high $1/B$

\end{abstract}

\section{Collective excitation spectrum of density modulated one dimensional
electron gas (DM1DEG)}

In the last decade or so remarkable progress has been made in epitaxial
crystal growth techniques which have made possible the fabrication of novel
semiconductor heterostructures. These modern microstructuring techniques can
be used to laterally confine quasi-two-dimensional electron gas (2DEG) in
e.g.,GaAs/AlGaAs heterostructure on a submicrometer scale to quasi-one
dimensional structures (quantum wires) or quasi-zero-dimensional quantum dots.
The system that we are considering in the present work can be realized by
various methods, e.g., the application of a laterally microstructured gate
electrode and the holographic illumination technique which allow for a tunable
periodic density modulation of the quasi-one-dimensional electron gas.

There have been a number of approaches to study collective excitations
spectrum of a quasi-two and one dimensional electron gas (1DEG) systems
theoretically \cite{1-5, 10, 11, 14, 26} and experimentally \cite{6-8, 12,
13}. We extend the theoretical work by calculating the inter- and intra Landau
band magnetoplasmon spectrum of a density modulated quasi-one dimensional
electron gas (DM1DEG) in the presence of a perpendicular magnetic field using
the Self-Consistent-Field (SCF) approach and focus on the oscillatory behavior
of the intra-Landau band magnetoplasmons. In this context, the term
\textquotedblleft1D\textquotedblright\ means that we start with the original
2DEG in $x-y$ plane. We apply a confining potential in the $x$-direction
leaving the $y$-direction free. A magnetic field is applied along the
$z$-direction perpendicular to the $x-y$ plane of the original 2DEG. The
collective particle energy spectrum, $E_{ij}(k_{y})=\frac{\hbar^{2}k_{y}^{2}%
}{2m^{\ast}}+E_{x}^{i}$ consists of energetically separated 1D subbands formed
due to lateral confining potential along the $x$-direction. The electron wave
vector $k_{y}$ characterizes the free motion in the $y$-direction.

The effect of density modulation is to broaden the Landau levels into
minibands whose width oscillates as a function of the magnetic field strength.
The electronic states are thus substantially altered, resulting in modulated
density of states, as shown by magnetocapacitance measurements \cite{12, 13}
of the quasi-two-dimensional systems. Behavior akin to this is expected for
the quasi-one dimensional system under consideration. The density of states
affects many response and transport phenomena as well as thermodynamic
properties. Of these, one of the most important properties is the collective
excitation spectrum and we evaluate the dynamic, nonlocal dielectric response
function to study it. As we show in our work for a quasi-one-dimensional
system, magnetoplasma spectrum significantly exhibits modulation of the
electronic density-of-states as oscillating magnetoplasma frequencies. As we
discuss in detail, this result is obtained in the regime of weak modulation
and long wavelength thus we don't find this result discussed in \cite{26}.
Another condition for the observation of these oscillations is that the
coupling between intra-Landau band mode and inter-Landau band mode must be
small. Mixing of these modes can be minimized by controlling the degree of
density modulation and by applying an appropriate magnetic field and
confinement potential.

\section{Formulation}

Our system is a density modulated quasi-One Dimensional Electron Gas (DM1DEG)
built on a two dimensional electron gas (2DEG) by inducing another effective
confining potential $\frac{1}{2}(m^{\ast}\omega_{0}^{2}x^{2}$) along the
$x$-direction which is assumed to be parabolic. The magnetic field is
perpendicular to the $x-y$ plane in which electrons with unmodulated areal
density $n_{D}$, effective mass $m^{\ast}$ and charge $-e$ are confined. We
employ the Landau gauge and write the vector potential as $A=(0,Bx,0)$. The
two-dimensional Schrodinger equation with parabolic confining potential in the
Landau gauge is ($\hbar=c=1$ here);
\begin{equation}
H_{0}=\frac{1}{2m^{\ast}}[-\frac{\partial^{2}}{\partial x^{2}}+(-i\frac
{\partial}{\partial y}+\frac{e}{c}Bx)^{2}]+\frac{1}{2}(m^{\ast}\omega_{0}%
^{2}x^{2}),\label{1}%
\end{equation}
Since the Hamiltonian does not depend on the $y$ coordinate, the unperturbed
wavefunctions are plane waves in the $y$-direction. This allows us to write
for the wavefunctions,
\begin{equation}
\phi_{nk_{y}}(\bar{x})=\frac{1}{\sqrt{L_{y}}}e^{ik_{y}y}u_{nk_{y}%
(x)},\label{2}%
\end{equation}
with $L_{y}$ being a normalization length in $y$-direction and $\bar{x}$ a 2D
position vector on the $x$-$y$ plane. Substitution of the above form of the
wavefunction in to equation (1), yields
\begin{equation}
H_{0}=-\frac{1}{2m^{\ast}}\frac{\partial^{2}}{\partial x^{2}}+\frac{1}%
{2}m^{\ast}\Omega_{0}^{2}(x-x_{0})^{2}+\frac{k_{y}^{2}}{2m^{\ast}%
(B)},\label{3}%
\end{equation}
with ($\Omega_{0}^{2}=\omega_{c}^{2}+\omega_{0}^{2},l^{2}(B)=l^{2}\frac
{\omega_{c}^{2}}{\Omega_{0}^{2}}$), where $\omega_{c}=\frac{eB}{m^{\ast}}$ is
the cyclotron frequency, $x_{_{0}}=-l^{2}(B)k_{y}=-\frac{\omega_{c}k_{y}%
}{m^{\ast}\Omega_{0}^{2}},$ is the coordinate of cyclotron orbit center and
$l(B)$ is the magnetic length, $m^{\ast}(B)=m^{\ast}\frac{\Omega_{0}^{2}%
}{\omega_{0}^{2}}$ is the normalized effective mass. In the $x$-direction, the
Hamiltonian has the form of a harmonic oscillator Hamiltonian. Hence we can
write the unmodulated eigenstates in the form $\phi_{nk_{y}}(\bar{x})=\frac
{1}{\sqrt{L_{y}}}e^{ik_{y}y}u_{nk_{y}(x;x_{0})}, $ with $u_{nk_{y}(x;x_{0}%
)}=(\sqrt{\pi}2^{n}n!l)^{\frac{-1}{2}}\exp(-\frac{1}{2l^{2}}(x-x_{0}%
)^{2})H_{n}(\frac{x-x_{0}}{l}),$ where $u_{nk_{y}(x;x_{0})} $ is a normalized
harmonic oscillator wavefunction centered at $x_{0}$ and $H_{n}(x)$ are
Hermite polynomials [15] with $n$ the Landau level quantum number. The energy
of $n$th Landau level (unperturbed Hamiltonian $H_{0}$) is
\begin{equation}
\varepsilon_{n}^{(0)}=(n+1/2)\Omega_{0}+\frac{k_{y}^{2}}{2m^{\ast}%
(B)}.\label{4}%
\end{equation}
In the presence of modulation, the Hamiltonian is augmented by the term
$H^{\prime}=V_{0}\cos(\frac{2\pi}{a}x),$ where $V_{0}$ is the amplitude of the
modulation and is about an order of magnitude smaller than Fermi energy
($V_{0}/\varepsilon_{F}$). Due to smallness of $V_{0}$ we employ first order
(in $H^{\prime}$) perturbation theory in the evaluation of the energy
eigenvalues using the unperturbed wavefunctions. The correction to the
unperturbed eigenenergies with a new variable $y=\frac{x-x_{0}}{l},$ is given by:%

\[
\varepsilon_{n}^{(1)}=\frac{2V_{0}}{\sqrt{\pi}2^{n}n!}\cos(\frac{2\pi}{a}%
x_{0})\underset{0}{\int}^{a}dy\exp(-y^{2})[H_{n}(y)]^{2}\cos(\frac{2\pi}%
{a}ly).
\]
The integral in above equation is given by Gradshteyn-Ryzhik [16] (page 841 \#
7.388.5); and the result is;
\begin{equation}
\varepsilon_{n}^{(1)}=V_{n}\cos(\frac{2\pi}{a}x_{0}),\label{5}%
\end{equation}
where $V_{n}=V_{0}\exp(-X/2)L_{n}(X),X=(\frac{2\pi}{a})^{2}\frac{\omega_{c}%
}{2m^{\ast}\Omega_{0}^{2}},$and $L_{n}(x)$ is a Laguerre polynomial. By
combining equations (4) and (5) we write, for the energy eigenvalues to first
order in $H^{\prime},$%
\begin{equation}
\varepsilon(n,x_{0})=(n+1/2)\Omega_{0}+\frac{k_{y}^{2}}{2m^{\ast}(B)}%
+V_{n}\cos(\frac{2\pi}{a}x_{0}).\label{6}%
\end{equation}
The above equation shows that the formerly sharp Landau levels, equation (4),
are now broadened into minibands by the modulation potential. Furthermore, the
Landau bandwidth (\symbol{126}$\mid V_{n}\mid$) oscillate as a function of
$n$, since $L_{n}(X)$ is an oscillatory function of its index \cite{15}.

\section{Density-Density Correlation Function of a DM1DEG in a Magnetic Field}

The dynamic and static response properties of an electron system are all
embodied in the structure of the density-density correlation function. We
employ the Ehrenreich-Cohen Self-Consistent Field (SCF) approach \cite{9} to
calculate the density-density correlation function. The SCF treatment
presented here is by its nature a high density approximation which has been
successful in the study of collective excitations in lower-dimensional systems
such as planar semiconductor superlattices \cite{17} and quantum wire
structures \cite{5, 8, 18}, both with and without an applied magnetic field.
Such success has been convincingly attested by the excellent agreement of SCF
predictions of plasmon spectra with experiments.

Following the SCF approach, the density response of electrons due to a
perturbing potential is given by%

\begin{align}
\delta n(\bar{x}_{0},z_{0};t)  & =\underset{\alpha\alpha^{\prime}}{\sum}%
\frac{f(\varepsilon_{\alpha^{\prime}})-f(\varepsilon_{\alpha})}{\varepsilon
_{\alpha^{\prime}}-\varepsilon_{\alpha}+\omega+i\eta}<\alpha\mid V(\bar
{x},z;\omega)\mid\alpha^{\prime}>\nonumber\\
\times & <\alpha^{\prime}\mid\delta(\bar{x}-\bar{x}_{0})\delta(z-z_{0}%
)\mid\alpha>,\label{7}%
\end{align}
where $V(\bar{x},z;\omega)$ is the self-consistent potential and $\alpha$
stands for the quantum numbers $n$ and $k_{y}$. Fourier transforming on the
$x-y$ plane we obtain the induced particle density
\begin{align*}
\delta n(\bar{q},z_{0};\omega)  & =\frac{1}{A}\delta(z)V(\bar{q}%
,z=0;\omega)\underset{\alpha\alpha^{\prime}}{\sum}\frac{f(\varepsilon
_{\alpha^{\prime}})-f(\varepsilon_{\alpha})}{\varepsilon_{\alpha^{\prime}%
}-\varepsilon_{\alpha}+\omega+i\eta}\\
\times & \mid<\underline{\alpha}^{\prime}\mid e^{-i\bar{q}.\bar{x}}%
\mid\underline{\alpha}>\mid^{2},
\end{align*}
where $A$ denotes the area of the system. We can perform the $k_{y}^{\prime}%
$-sum in the above equation to obtain
\begin{align}
\delta n(\bar{q},z;\omega)  & =\frac{1}{A}\delta(z)V(\bar{q},z=0;\omega
)\nonumber\\
& \times\underset{n,n^{\prime},k_{y}}{\sum}C_{nn^{\prime}}(\frac{\bar{q}%
^{2}\omega_{c}}{2m^{\ast}\Omega_{0}^{2}})\frac{f(\varepsilon(n^{\prime}%
,k_{y}-q_{y}))-f(\varepsilon(n,k_{y}))}{\varepsilon(n^{\prime},k_{y}%
-q_{y})-\varepsilon(n,k_{y})+\omega+i\eta}.\label{8}%
\end{align}
Writing the induced particle density as $\delta n(\bar{q},z;\omega)=\delta
n(\bar{q},\omega)\delta(z),$ allows us to rewrite equation (8) as $\delta
n(\bar{q},\omega)=V(\bar{q},\omega)\Pi_{0}(\bar{q},\omega)$: where $V(\bar
{q},\omega)=V(\bar{q},z=0;\omega)$ and $\Pi_{0}(\bar{q},\omega)$ is the
density-density correlation function of the non-interacting electron system,
given by
\begin{equation}
\Pi_{0}(\bar{q},\omega)=\frac{1}{A}\underset{n,n^{\prime}}{\sum}%
\underset{k_{y}}{\sum}C_{nn^{\prime}}(\frac{\bar{q}^{2}\omega_{c}}{2m^{\ast
}\Omega_{0}^{2}})\frac{f(\varepsilon(n^{\prime},k_{y}-q_{y}))-f(\varepsilon
(n,k_{y}))}{\varepsilon(n^{\prime},k_{y}-q_{y})-\varepsilon(n,k_{y}%
)+\omega+i\eta},\label{9}%
\end{equation}
where
\[
C_{nn^{\prime}}(x)=\frac{n_{2}!}{n_{1}^{\prime}!}e^{-x}x^{n_{1}-n_{2}%
}[L_{n_{2}}^{n_{1}-n_{2}}(x)]^{2}%
\]
with $n_{1}$= max($n$, $n^{\prime}$), $n_{2}$= min($n$, $n^{\prime}$), and
$L_{n}^{\prime}(x)$ an associated Laguerre polynomial. The induced potential
$V^{ind}$ is related to the density response $\delta n$ by Poisson's equation
\begin{equation}
\nabla^{2}V^{ind}(\bar{x},z;t)=-\frac{4\pi e^{2}}{k}\delta n(\bar
{x},z;t),\label{10}%
\end{equation}
where $k$ is the background dielectric constant. The above equation can be
solved to yield
\begin{equation}
V^{ind}(\bar{q},\omega)=\frac{2\pi e^{2}}{k\bar{q}}\delta n(\bar{q}%
,\omega).\label{11}%
\end{equation}
Recalling that the self-consistent potential, $V(\bar{q},\omega)=V^{ext}%
(\bar{q},\omega)+V^{ind}(\bar{q},\omega),$ is the sum of the external and
induced potentials, multiplying both sides by $\Pi(\bar{q},\omega)$ and
solving for $\delta n(\bar{q},\omega)$ yields
\begin{equation}
\delta n(\bar{q},\omega)=\Pi(\bar{q},\omega)V^{ext}(\bar{q},\omega),\label{12}%
\end{equation}
where
\begin{equation}
\Pi(\bar{q},\omega)=\frac{\Pi_{0}(\bar{q},\omega)}{1-v_{c}(\bar{q})\Pi
_{0}(\bar{q},\omega)}\label{13}%
\end{equation}
is the density-density correlation function of the interacting system with
$v_{c}(\bar{q})=$ $\frac{2\pi e^{2}}{k\bar{q}}$ the 2-D Coulomb potential.
Making use of the transformation $k_{y}\rightarrow-k_{y}$ with the fact that
$\varepsilon(n,k_{y})$ is an even function of $k_{y},$ and at the same time
interchanging $n\leftrightarrow n^{\prime}$ we write for the non-interacting
density-density correlation function equation (8)
\begin{align}
\Pi_{0}(\bar{q},\omega)  & =\frac{m^{\ast}\Omega_{0}^{2}}{\pi a\omega_{c}}\sum
C_{nn^{\prime}}(\frac{\bar{q}^{2}\omega_{c}}{2m^{\ast}\Omega_{0}^{2}}%
)\overset{a}{\underset{0}{\int}}dx_{0}[f(\varepsilon(n,x_{0}+x_{0}^{\prime
})-f(\varepsilon(n^{\prime},x_{0}))]\nonumber\\
& \times\lbrack\varepsilon(n,x_{0}+x_{0}^{\prime})-\varepsilon(n^{\prime
},x_{0})+\omega+i\eta]^{-1}.\label{14}%
\end{align}
In writing the above equation we converted the $k_{y}$-sum into an integral
over $x_{0}$. $f(E)$ is the Fermi -Dirac distribution function, $x_{0}%
=-\frac{k_{y}\omega_{c}}{m^{\ast}\Omega_{0}^{2}}$ and $x_{0}^{\prime}%
=-\frac{q_{y}\omega_{c}}{m^{\ast}\Omega_{0}^{2}}.$

The above equations (13,14) will be the starting point of our examination of
the inter-and intra-Landau band plasmons. The form of the expressions for the
real and imaginary part of the density-density correlation function makes the
even and odd (in frequency $\omega$) properties of these functions very
apparent. These functions are the essential ingredients for theoretical
considerations of such diverse problems as high frequency and steady state
transport, static and dynamic screening and correlation phenomena.

The plasma modes are readily furnished by the singularities of the function
$\Pi(\bar{q},\omega),$ from the roots of the longitudinal plasmon dispersion
relation obtained from equation (13) as
\begin{equation}
1-v_{c}(\bar{q})\operatorname{Re}\Pi_{0}(\bar{q},\omega)=0\label{15}%
\end{equation}
along with the condition $Im\Pi_{0}(\bar{q},\omega)=0$ to ensure long-lived
excitations. The roots of equation (15) give the plasma modes of the system.
\begin{equation}
1=\frac{2\pi e^{2}}{k\bar{q}}\frac{m^{\ast}\Omega_{0}^{2}}{\pi a\omega_{c}%
}\underset{n,n^{\prime}}{\sum}C_{nn^{\prime}}(\frac{\bar{q}^{2}\omega_{c}%
}{2m^{\ast}\Omega_{0}^{2}})(I(\omega)+I(-\omega)),\label{16}%
\end{equation}
with
\[
I(\omega)=P\int\limits_{0}^{a}dx_{0}\frac{f(\varepsilon(n,x_{0}))}%
{+\omega-\varepsilon(n,x_{0})+\varepsilon(n^{\prime},x_{0}+x_{0}^{\prime})},
\]
where $P$ is the principal value.

The plasma modes originate from two kinds of electronic transitions, those
involving different Landau bands (inter-Landau band plasmons) and those within
a single Landau-band (intra-Landau band plasmons). Inter-Landau band plasmons
involve the local 1D magnetoplasma mode and the Bernstein-like plasma
resonances \cite{20}, all of which involve excitation frequencies greater than
the Landau-band separation ($\sim\Omega_{0}$). On the other hand, intra-Landau
band plasmons resonate at frequencies comparable to the bandwidths, and the
existence of this new class of modes is due to finite width of the Landau
levels. These magnetoplasmons in quasi-one-dimensional system have been
analysed in detail elsewhere \cite{26}. We will concentrate on the oscillatory
behavior of these magnetoplasmons. The occurrence of such intra-Landau band
plasmons is accompanied by SdH type of oscillatory behavior \cite{2, 21} in
$1/B$ . These oscillations \cite{2, 22} are not with constant period in $1/B$
(which exhibits significant effect for small value of $B$ and corresponding
large value of $1/B$) due to confinement potential acting in the $x$-direction
and also show the depopulation, and cross over effects \cite{14} on
magnetoplasmons from density modulated two-dimensional electron gas (DM2DEG)
\cite{23} to density modulated one-dimensional electron gas (DM1DEG).

SdH type of oscillations result from the emptying out of electrons from
successive Landau bands when they pass through the Fermi level as the magnetic
field is increased. The amplitude of the SdH type of oscillations is a
monotonic function of magnetic field, when the Landau bandwidth is independent
of the band index $n$. In DM1DEG considered here, the Landau bandwidths
oscillate as a function of the band index $n.$ It is to be expected that such
oscillating bandwidths would effect the plasmon spectrum of the intra-Landau
band type, resulting in another type of oscillation. These oscillations are
not with constant period in $1/B,$ because at small value of $B$ and
corresponding large value of $1/B$ cyclotron diameter exceeds the
characteristic length of the confining potential.

For the excitation spectrum, we need to numerically solve equation (15) for
all vectors, frequencies, magnetic field, and confinement potential. We will
consider the case of weak modulation ($V_{0}/E_{F}<<1$) and long wave length.
In these limits we can solve equation (16) analytically for zero temperature.
We expand the coefficient $C_{nn^{\prime}}(\frac{\bar{q}^{2}\omega_{c}%
}{2m\Omega_{0}^{2}})$ to lowest order in its argument with the result
\begin{equation}
1=\frac{2\pi e^{2}}{km^{\ast}}\bar{q}\frac{1}{\omega^{2}-\Omega_{0}^{2}}%
(\frac{m\Omega_{0}}{\pi a}\underset{n}{\sum}f(\varepsilon_{n}))\label{17}%
\end{equation}
The term in parentheses is easily recognized as the unmodulated particle
density $n_{D}=\frac{m\Omega_{0}}{\pi a}\underset{n}{\sum}f(\varepsilon_{n}),$
where $n$ is sum over all occupied Landau bands. Defining the plasma frequency
through $\omega_{p,D}^{2}=\frac{2\pi n_{D}e^{2}}{km}\bar{q},$we finally obtain
the inter-Landau-band plasmon dispersion relation $1=\frac{\omega_{p,D}^{2}%
}{\omega^{2}-\Omega_{0}^{2}}$ or $\omega^{2}=\Omega_{0}^{2}+\omega_{p,D}^{2},$
which is the ordinary one-dimensional plasmon dispersion relation.

The intra-Landau-band plasmon dispersion relation for zero temperature reduces
to $1=\frac{\overset{\sim}{\omega}^{2}}{\omega^{2}}$, where
\begin{equation}
\overset{\sim}{\omega}^{2}=\frac{8e^{2}}{k\bar{q}}\frac{m^{\ast}\Omega_{0}%
^{2}}{\pi a\omega_{c}}\sin^{2}(\frac{\pi}{a}(x_{0}^{\prime})\times\underset
{n}{\sum\mid}V_{n}\mid\sqrt{1-\Delta_{n}^{2}}\theta(1-\Delta_{n}),\label{18}%
\end{equation}
with $A_{n}=\frac{a}{2\pi}\frac{\mid V_{n}\mid}{V_{n}}\sqrt{1-\Delta_{n}^{2}%
}\theta(1-\Delta_{n}),\Delta_{n}=\mid\frac{\varepsilon_{F}-\varepsilon_{n}%
}{V_{n}}\mid,\theta(x)$ the Heaviside unit step function.

We have derived the expression for $\overset{\sim}{\omega}$ (equation 18)
under the condition $\omega>>\mid\varepsilon(n,x_{0}+x_{0}^{\prime
})-\varepsilon(n,x_{0})\mid$ as $x_{0}^{\prime}\rightarrow0$ which leads to a
relation between the frequency and the Landau level broadening $\omega
>>\mid2V_{n}\sin(\frac{\pi}{a}x_{0}^{\prime})\sin[(\frac{2\pi}{a})(x_{0}%
+\frac{x_{0}^{\prime}}{2})]\mid$. This ensures that $Im\Pi_{0}(\bar{q}%
,\omega)=0$ and the intra-Landau-band magnetoplasmons are undamped. For a
given $V_{n}$, this can be achieved with a small but nonzero $q_{y}$ (recall
that $x_{0}^{\prime}=-\frac{q_{y}\omega_{c}}{m^{\ast}\Omega_{0}^{2}}$).

In general, the inter- and intra-Landau-band modes are coupled for arbitrary
magnetic field strengths. The general dispersion relation is :
\[
1=\frac{\omega_{p,D}^{2}}{\omega^{2}-\Omega_{0}^{2}}+\frac{\overset{\sim
}{\omega}^{2}}{\omega^{2}}
\]
This equation yields two modes which are given by :
\begin{align*}
\omega_{\pm}^{2}  & =\frac{1}{2}(\Omega_{0}^{2}+\omega_{p,D}^{2}+\overset
{\sim}{\omega}^{2})\pm\frac{1}{2}\{(\Omega_{0}^{2}+\omega_{p,D}^{2}%
+\overset{\sim}{\omega}^{2}+2\Omega_{0}\overset{\sim}{\omega})\\
& \times(\Omega_{0}^{2}+\omega_{p,D}^{2}+\overset{\sim}{\omega}^{2}%
-2\Omega_{0}\overset{\sim}{\omega})\}^{1/2}%
\end{align*}
which reduces to:
\[
\omega_{+}^{2}=\Omega_{0}^{2}+\omega_{p,D}^{2},
\]
and
\[
\omega_{-}^{2}=\overset{\sim}{\omega}^{2}
\]
with corrections of order $\overset{\sim}{\omega}^{2}/\Omega_{0}^{2}$ and
$\overset{\sim}{\omega}^{2}/\omega_{p,D}^{2}$ . So long as $\mid V_{n}%
\mid<\Omega_{0},$ mixing of the inter-and intra-band modes is small. Only the
intra-Landau-band mode ( $\overset{\sim}{\omega})$ will be excited in the
frequency regime $\Omega_{0}>\omega\sim$ $\mid V_{n}\mid.$ We now present the
results for the oscillatory bahavior of the intra-Landau band magnetoplasmons
as a function of $1/B$ and the confinement energy.

\section{ \ Numerical Results and Discussion}

The intra-Landau-band plasma frequency given by equation (18) is shown
graphically in Figure (1) as a function of $1/B$ for two different values of
confinement energy, using parameters \cite{2, 5, 7}: $m^{\ast}=0.07m_{e}$,
$k=13.6$, $n_{D}=6\times10^{15}$ m$^{-2}$, $a=500$nm, and $V_{0}$ = 1.0 meV;
also we take $q_{x}=0$ and $q_{y}=0.01k_{F},$ with $k_{F}=(2\pi n_{D})^{1/2}$
being the Fermi wave number of the unmodulated 1DEG in the absence of magnetic
field. The modulation induced oscillations are apparent, superimposed on
SdH-type oscillations. These oscillations are not with constant period in
$1/B$ (which exhibits significant effect for small value of $B$ and
corresponding large value of $1/B$) due to confinement potential acting in the
$x$-direction. They have longer period and much reduced amplitude. It must be
noted that this result is obtained in the regime of weak modulation and long
wavelength. Another condition for the observation of these oscillations is
that the coupling between inter-Landau band mode and intra-Landau band mode
must be small. These modes involve different energy scales $\omega>\Omega_{0}$
for the former and $\omega\sim\mid V_{n}\mid<\Omega_{0}$ for the latter.
Mixing of these modes can be minimized by controlling the degree of density
modulation and by applying an appropriate magnetic field and confinement potential.

The origin of two types of oscillations can be understood by a closer analytic
examination of equation (18). In the regime, $\Omega_{0}>\left\vert
V_{n}\right\vert $, the unit step function vanishes for all but the highest
occupied Landau band, corresponding , say , to the band index $N$. The sum
over $n$ is trivial, and plasma frequency is given as $\overset{\sim}{\omega
}^{2}=\mid V_{N}\mid^{1/2}(1-\Delta_{N}^{2})^{1/4}\theta(1-\Delta_{N})$. The
analytic structure primarily responsible for the SdH type of oscillations is
the function $\theta(1-\Delta_{N})$, which jumps periodically from zero (when
the Fermi level is above the highest occupied Landau band) to unity (when the
Fermi level is contained with in the highest occupied Landau band). On the
other hand, the periodic modulation of the amplitude of the SdH type
oscillations shown in Figures (1 \& 2) is due to the oscillatory nature of the
factor $\mid V_{N}\mid^{1/2}$, which has been shown in a two-dimensional
system to exhibit commensurability oscillations \cite{6, 12, 22, 24}. In our
case, these oscillations are not with constant period in $1/B.$ This clearly
indicates the one dimensional character of our theory. In a DM2DEG the number
of occupied Landau levels increases with decreasing $B$, leading, ideally, to
an infinite number of SdH type of oscillations periodic in $1/B$ \cite{23}. In
a 1DEG system however, only a finite number of 1D subbands are occupied at
$B=0$, giving rise to finite number of SdH type of oscillations and deviations
from the $1/B$ period, because with increasing $B$ the 1D density of states
increases and the hybrid 1D subband Landau levels are depopulated \cite{4, 13,
25}. In the extreme 2D regime ($\omega_{0}<<\omega_{c}$), the Fermi energy
goes to the bottom of the 1D Landau subband. If we lower the confinement
potential the magnetic confinement overcomes the potential confinement, hence
we are in the original 2D regime \cite{23}. On the other hand if we increase
the confinement potential, the confinement potential overcomes the magnetic
confinement and we have a crossover \cite{14} from a two-dimensional system to
a one-dimensional system. In figures (1, 2) we have plotted the intra-Landau
band plasma frequency as a function of the inverse magnetic field \ for two
different confinement energies given by Eq. (18).

\section{Conclusion}

We have determined the intra-Landau band plasmon frequency for a density
modulated quasi-one dimensional electron gas in the presence of a magnetic
field employing the SCF approach . Furthermore, we have seen the oscillations
of the intra-Landau band plasma frequency in 1D regime as a function of
$B^{-1}$, their origin lies in the interplay of the three physical length
scales of the system i.e. the modulation period, confinement length and
cyclotron diameter at the Fermi level. When a strong magnetic field is
applied, our model recovers complete Landau quantization and for a very high
magnetic field our results are comparable with extreme 2D regime ($\Omega
_{0}<<\omega_{c}$).

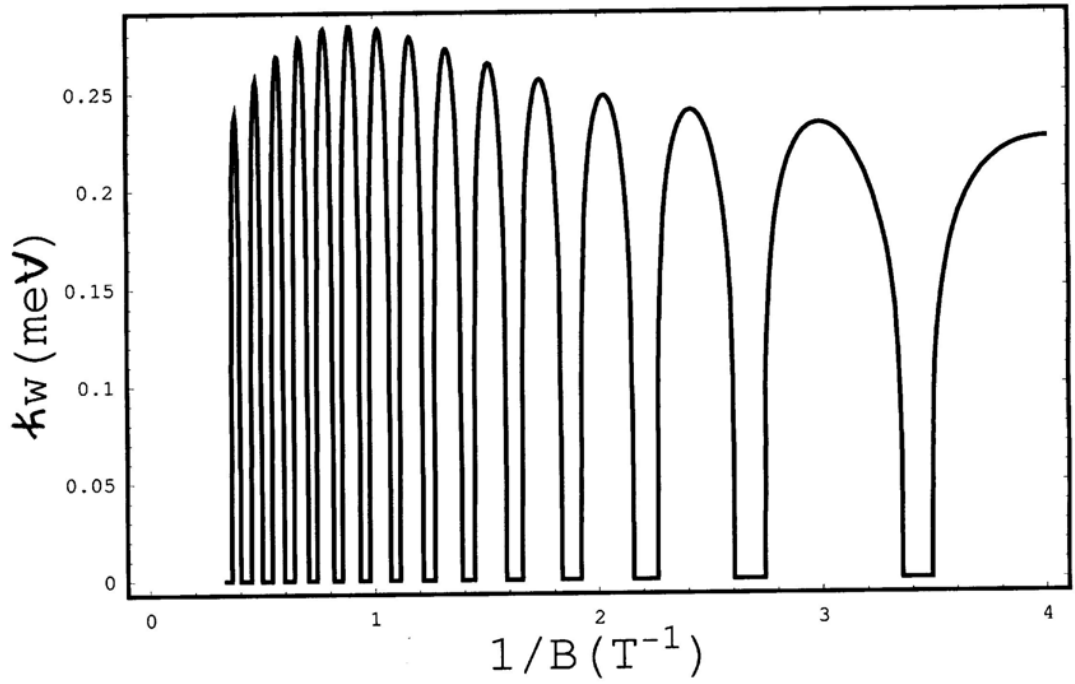
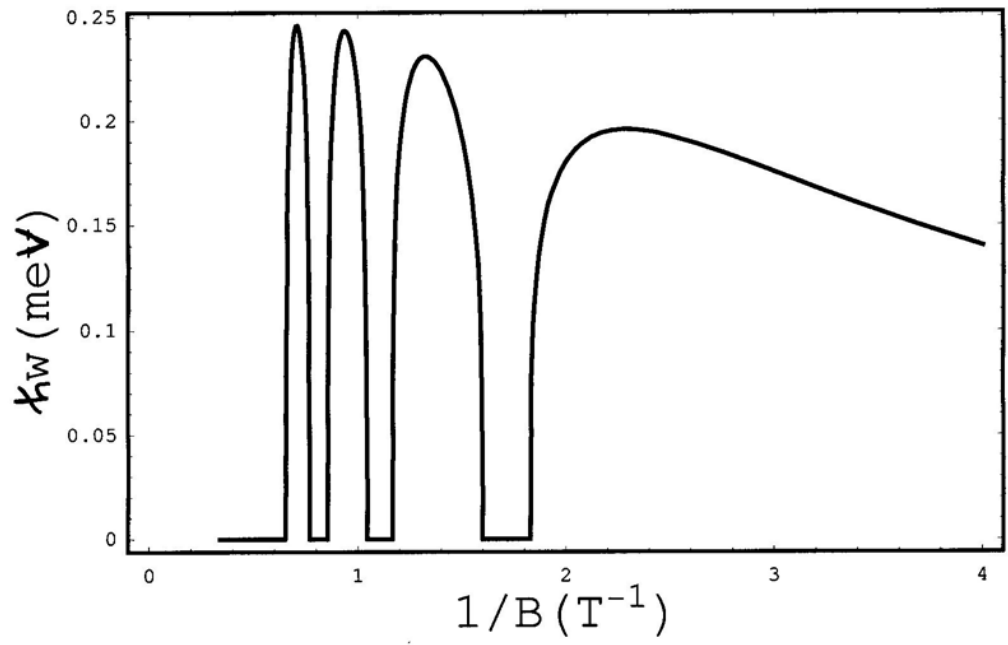
\end{document}